\documentclass{article}
\usepackage{spconf,amsmath,graphicx}
\usepackage{subfiles} 
\usepackage{multirow}
\usepackage{booktabs}

\def\L{{\cal L}}

\title{Poformer: A simple pooling transformer for speaker verification}
\name{Yufeng Ma$^1$\textsuperscript{*}, Yiwei Ding$^{1,2}$\textsuperscript{*}, Miao Zhao$^1$, Yu Zheng$^1$, Min Liu$^1$, Minqiang Xu$^1{}^\dag$}
\address{$^1$ SpeakIn Technologies Co. Ltd. \\ $^2$ Fudan University}
\begin{document}
%
\maketitle

\begingroup\renewcommand\thefootnote{*}
\footnotetext{Equal contribution}
\begingroup\renewcommand\thefootnote{\dag}
\footnotetext{Corresponding author}
\begin{abstract}
Most recent speaker verification systems are based on the extraction of speaker embeddings using a deep neural network. The pooling layer in the network aims to aggregate frame-level features extracted by the backbone. In this paper, we propose a new transformer based pooling structure called PoFormer to enhance the ability of the pooling layer to capture information along the whole time axis. Different from previous works that apply attention mechanism in a simple way or implement the multi-head mechanism in serial instead of in parallel, PoFormer follows the initial transformer structure with some minor modifications like a  positional encoding generator, drop path and LayerScale to make the training procedure more stable and to prevent overfitting. Evaluated on various datasets, PoFormer outperforms the existing pooling system with at least a 13.00\% improvement in EER and a 9.12\% improvement in minDCF.
\end{abstract}
\begin{keywords}
speaker verification, transformer
\end{keywords}
\section{Introduction}
\label{sec:intro}
Speaker verification aims to determine whether two audio segments belong to the same speaker. This can be done by embedding the audio segments into high-dimensional vectors and then measuring their similarity. Before deep learning is applied in speaker verification, methods like i-vectors \cite{ivector} achieved a good performance. In recent years, progresses have been made by DNN-based systems, such as DNN-embeddings \cite{dnn_17_interspeech}, x-vectors \cite{x-vector} and ECAPA-TDNN \cite{ecapa_first}. In most systems, the whole network can be divided into three parts: (1) a backbone to extract the features of the audio segment, (2) a pooling layer to capture global information in the extracted feature and (3) an embedding layer with a loss function to classify the speaker. Our work focuses on the pooling layer.

In previous works, pooling layers like statistical pooling \cite{x-vector} are proposed and achieve good results despite their simplicity. Moreover, introducing attention-based modules has become popular in speaker verification recently. Works like attentive statistical pooling \cite{okabe_2018_attentive}, cas pooling \cite{ecapa_first}, self-attentive speaker embedding \cite{Zhu2018SelfAttentiveSE} and serialized multi-layer multi-head attention \cite{zhu_2021_serialized} prove that attention mechanism is effective in aggregating frame-level features.

Transformer, first proposed in \cite{google_2017_attention}, is also an attention-based structure and has made a lot of successes in various areas including natural language processing \cite{google_2018_bert} and computer vision \cite{dosovitskiy_2021_vit, carion2020endtoend, zheng2021rethinking}, inspired by which we propose a pooling transformer, PoFormer by introducing transformer to the pooling layer of our speaker verification network to strengthen its capability of capturing information across the whole time domain. Different from \cite{zhu_2021_serialized} which re-designs the inner structure of the attention module and implements the multi-head mechanism serially, we strictly follows \cite{google_2017_attention} where different heads are in parallel, providing a simple but effective pooling transformer.

ViT \cite{dosovitskiy_2021_vit} sets all the positional encoding as learnable. However, it causes a performance decay on input sequences with various length. To make our model robust to audio segments with different length, we use positional encoding generator (PEG) \cite{chu2021conditional} to generate the positional encoding for the input of the transformer. Following \cite{chu2021conditional}, we use multiple PEGs, one for each transformer layer.

Furthermore, because of the complexity of transformer, it is vulnerable to overfitting. Therefore, we add LayerScale \cite{touvron2021going} and drop path \cite{larsson2017fractalnet} in our PoFormer. The performance is evaluated on Voxceleb1-O, Voxceleb1-E and Voxceleb1-H \cite{Nagrani_2017_voxceleb, Chung_2018_voxceleb2}. With PEG, LayerScale and drop path, our model achieves a good result. Evaluated on the hardest Voxceleb1-H dataset, the equal error rate (EER) is 13.00\% better than the baseline and the minimum decision cost function (minDCF) is 9.12\% better than the baseline.

Two main contributions of our paper are: 1) we first introduce the original transformer structure into the pooling layer of a speaker verification network and 2) we provide a set of effective strategies to improve the performance of our PoFormer and, at the same time, make it more robust.

\section{Proposed Methods}
\label{sec:methods}
\begin{figure}[ht]
  \centering
  \includegraphics[width=0.48\textwidth]{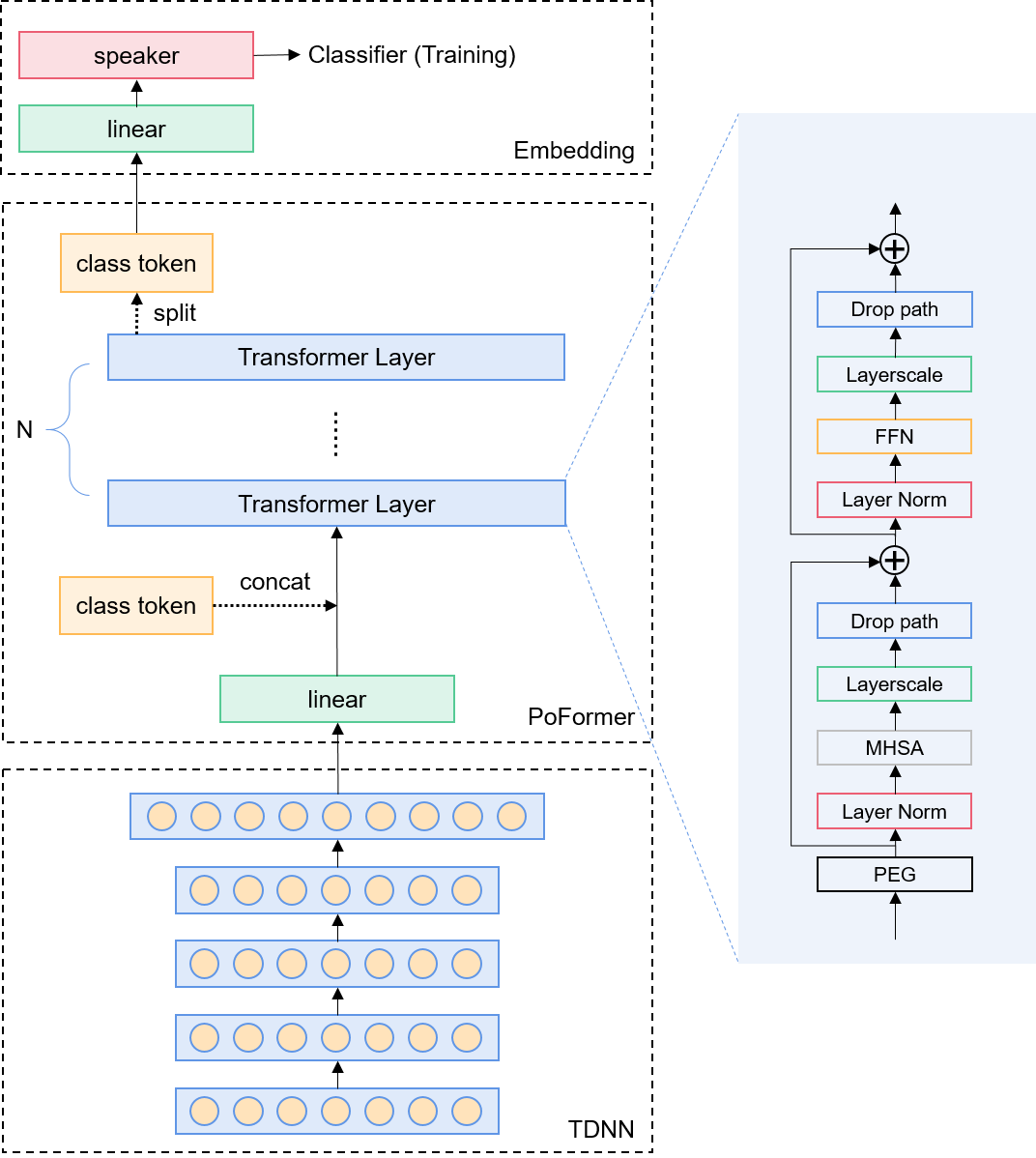}
  \caption{The overall structure of our speaker verification system.}
  \label{fig:overall}
\end{figure}

As the pooling layer aims to aggregate frame-level features and the transformer is known of its strong capability of capturing global information, we propose a transformer-based pooling layer, named as PoFormer (pooling transformer). This section first describes the overall structure of our speaker verification network and then focuses on the PoFormer.

The overall structure of our proposed system is shown in \textbf{Fig.\,\ref{fig:overall}}. We follow the popular design where the whole network is divided into three parts: a backbone, a pooling layer and an embedding layer. In the backbone part, 5 TDNN layers \cite{hinton1989tdnn} are used to extract the frame-level features of the utterance. Then the features are fed into our PoFormer to integrate the feature sequence into one vector. After that, an embedding layer is used to obtain the speaker embedding. The network is trained as a speaker classifier with an AM softmax loss \cite{Wang_2018_amsoftmax}. For comparison, we also construct a baseline system where the PoFormer is substituted by a simple statistic pooling \cite{x-vector}.

\subsection{PoFormer Architecture}
The stucture of PoFormer is also illustrated in \textbf{Fig.\,\ref{fig:overall}}. The dimension of TDNN's output is usually high, making the model computationally intensive as well as memory requiring. To reduce the cost of space and time, the frame-level features are first compressed in dimension by a linear layer, and then fed into several stacked transformer layers.

\textbf{Multi-head self-attention (MHSA)} module is the core of the transformer. Given an input sequence $X\in R^{L \times d}$ where $L$ denotes the sequence length and $d$ denotes the feature dimension, $Q_i\in R^{L \times d_k}$, $K_i\in^{L\times d_k}$ and $V_i\in R^{L\times d_v}$ where $i$ denotes the $i$-th head are first generated by learnable linear projections $Q_i=XW_i^Q$, $K_i=XW_i^K$ and $V_i=XW_i^V$. Then the output from the $i$-th head can be formulated as
$$Attn_i(X) = Softmax(\dfrac{Q_iK_i^T}{\sqrt{d_k}})V_i,\quad i=1, 2, ..., n$$
where $n$ is the number of heads. After that, the outputs from different heads are concatenated together and transformed into the final output of the MHSA module by a linear layer:
$$MHSA(X) = Concat([Attn_1, Attn_2, ..., Attn_n])W^O$$
where $W^O\in R^{nd_k\times d}$ is also a learnable parameter.

\textbf{Feedforward net (FFN)} module is also important in transformer. It consists of two linear layers with one non-linear layer in between. Formally, given the input $X$, the output can be written as
$$FFN(X) = f(XW_1 + b_1)W_2 + b_2$$
where $f$ denotes a non-linear activation function such as ReLU or GELU, and $W_1$, $W_2$, $b_1$ and $b_2$ are the parameters of two linear layers.

To make the transformer easier to train, layer normalization and residual connection are used and the layer normalization is added before the MHSA and the FFN module. The output from the $i$-th transformer layer can be formulated as:
\begin{align}
   \label{eq:mhsa_pre} \tilde{X}^{(i)} &= X^{(i)} + MHSA(LN(X^{(i)}))\\
   \label{eq:ffn_pre} X^{(i+1)} &= \tilde{X}^{(i)} + FFN(LN(\tilde{X}^{(i)}))
\end{align}
where $LN(\cdot)$ denotes layer normalization. Several layers are stacked together to strengthen the capability of the model. Moreover, following ViT, a class token is appended to the feature sequence and used in later embedding and classification.

\begin{table*}[t]
  \centering
  \caption{Performances of our proposed PoFormer and the comparison with statistical pooling and serialized multi-head multi-layer pooling. All the models are based on our own implementation and are trained with the same 1024-dimensional TDNN backbone.}
  \label{tab:tabel_one}
  \setlength{\tabcolsep}{3.82mm}{
  \begin{tabular}{lcccccc}
    \toprule
    \multirow{4}{*}{Models} & 
    \multicolumn{2}{c}{\multirow{2}{*}{\textbf{VoxCeleb1-O}}} & 
    \multicolumn{2}{c}{\multirow{2}{*}{\textbf{VoxCeleb1-E}}} &  \multicolumn{2}{c}{\multirow{2}{*}{\textbf{VoxCeleb1-H}}} \\ \\
    \cline{2-7} 
    & \multirow{2}{*}[-2pt]{\textbf{EER(\%)}} &  \multirow{2}{*}[-2pt]{$\textbf{minDCF}_\textbf{0.01}$} & \multirow{2}{*}[-2pt]{\textbf{EER(\%)}} & \multirow{2}{*}[-2pt]{$\textbf{minDCF}_\textbf{0.01}$} & 
     \multirow{2}{*}[-2pt]{\textbf{EER(\%)}} & \multirow{2}{*}[-2pt]{$\textbf{minDCF}_\textbf{0.01}$} \\ \\
    \midrule
    Snyder et al. (baseline) \cite{x-vector}  & 2.0943 & 0.1834   & 1.9461 & 0.1958   & 3.1885 & 0.2773 \\
    Zhu et al. \cite{zhu_2021_serialized} & 1.8558 & 0.2526   & 1.9874 & 0.2184   & 3.4930 & 0.3207 \\
    \midrule
    PoFormer ($N = 3$)  & 1.6172 & 0.1499   & 1.6916 & 0.1756   & 2.8754 & 0.2572 \\
    PoFormer ($N = 5$)  & 1.5641 & 0.1345   & 1.6547 & 0.1698   & 2.8255 & 0.2593 \\
    PoFormer ($N = 7$)  & \textbf{1.5218} & \textbf{0.1298}   & \textbf{1.6203} & \textbf{0.1661}   & \textbf{2.7739} & \textbf{0.2520}\\

    \bottomrule
  \end{tabular}}
  \label{tab:a0}
  
\end{table*}

\subsection{Positional Encoding Generator}

\begin{figure}[ht]
  \centering
  \includegraphics[width=0.3\textwidth]{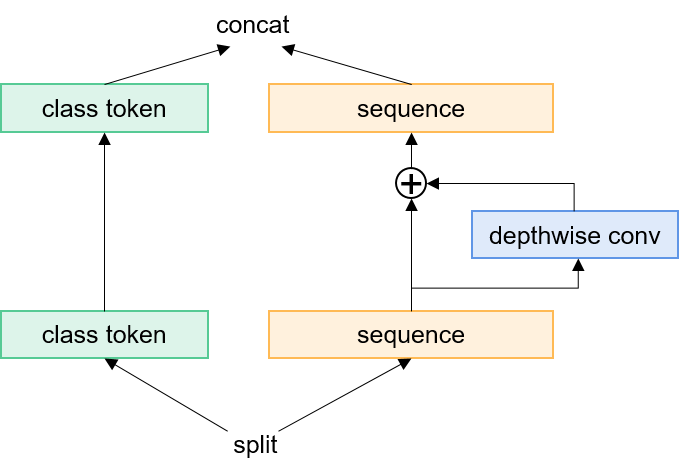}
  \caption{The structure of positional encoding generator (PEG). Note that the class token is first split and concatenated back after the position encoding is generated.}
  \label{fig:peg}
\end{figure} 

To make the input of transformer permutation-variant but translation-invariant, we use a 1-D depth-wise convolution layer to generate the positional encoding. The class token isn't involved in the generation of positional encoding. That is, as is illustrated in \textbf{Fig.\,\ref{fig:peg}}, before calculating positional encoding, the class token is split away from the feature. After the generated positional encoding is added to the frame-level features, the class token is concatenated back and fed into the MHSA module. Moreover, we apply one PEG module before every transformer layer to improve the performance.

\subsection{Drop path and LayerScale}
To make the training process more stable and to avoid overfitting, we add LayerScale and drop path into our PoFormer. 

\textbf{LayerScale} means scaling the output of the MHSA module and the FFN module in a channel-wise way with the scale factors learnable. The technique helps to stabilize the training in early stages. Formally, let $\gamma^{(i)} \in R^{d}$ denote the scale factor, then LayerScale transforms equation \ref{eq:mhsa_pre} and \ref{eq:ffn_pre} into
\begin{align}
   \tilde{X}^{(i)} &= X^{(i)} + MHSA(LN(X^{(i)}))\times diag(\gamma_1^{(i)})\\
   X^{(i+1)} &= \tilde{X}^{(i)} + FFN(LN(\tilde{X}^{(i)}))\times diag(\gamma_2^{(i)})
\end{align}

\textbf{Drop path} is inserted after the MHSA or FFN module before its output is added to residual connection branch. It sets the output of the MHSA or FFN module as zero with a certain probability, so that the input is passed forward by the residual connection with no changes made. This prevents the model from overfitting the training data.

\section{Experiments}
\label{sec:experiments}
\subsection{Dataset}
We used VoxCeleb2 development set \cite{Chung_2018_voxceleb2} to train all of our models and the data augmentation techniques were the same as in \cite{zhao2021speakin} except that we didn't apply the online data augmentation. After augmentation, we had 16,380,135 audio segments from 17,982 speakers. The models were evaluated by their equal error rate (EER) and minimum decision cost funcion (minDCF) on VoxCeleb1-O, VoxCeleb1-E and VoxCeleb1-H \cite{Nagrani_2017_voxceleb, Chung_2018_voxceleb2}.

\subsection{Experiment and Results}
We used a 1024-dimensional TDNN backbone and the output dimension was 1500. Both the dimension of PoFormer and the speaker embedding were set as 512. The number of heads in PoFormer was 4 and the dimension of the feed forward layer was 1024, twice the PoFormer dimension. Drop path rate was modified according to the number of layers. For Poformers with $3$, $5$ and $7$ transformer layers, the drop path rate was set as $0.3$, $0.4$ and $0.45$ respectively.

We used 81-dimensional fbank feature with no extra voice activation detection as the input. During training, 300 frames were extracted for each audio segment. The margin and the scale of AMSoftmax were 30 and 0.25 respectively. Our system was trained using an AdamW optimizer with a weight decay of 0.2. A cosine annealing learning rate scheduler was applied with an initial learning rate of 1e-3 and a minimum learning rate of 5e-5, and the training procedure lasted for 100,000 steps. Furthermore, a 10,000-step learning rate warm-up was used for the sake of stability. 

\textbf{Table \ref{tab:a0}} reports the performance of PoFormers with different number of layers on three test datasets. We can see that our PoFormer outperforms both the baseline and serialized multi-head multi-layer attention module (reproduced by us using the same backbone and embedding layer). Our 7-layer PoFormer improved the EER by 13.00\% and the minDCF by 9.12\% compared with the baseline.

\subsection{Ablation Studies}

\subsubsection{Drop path}

To show the importance of drop path, we present the performances of PoFormers with different drop path rate in \textbf{Table \ref{tab:droppath}}. Only results on the most difficult test dataset Voxceleb1-H are listed (the same below). As is shown in the table, the performance is poor when there is no drop path, and the best result is observed with a drop path rate of 0.3. The PoFormers consist of three transformer layers and are trained without positional encoding. Note that as the drop path rate can be sensitive to the number of transformer layers, we recommend a larger drop path rate for more transformer layers. 

\begin{table}[htbp]
  \centering
  \caption{Ablation study on drop path rate. PoFormers trained with different drop path rate and evaluated on Voxceleb1-H are presented.}
  \label{tab:tabel_two}
  \setlength{\tabcolsep}{5.4mm}{
  \begin{tabular}{lcc}
    \toprule
    {Drop path rate $p$} & {\textbf{EER(\%)}} & {$\textbf{DCF}_\textbf{0.01}$} \\
    \midrule
    $p=0$ (no drop path)      & 6.0909 & 0.5043 \\
    $p=0.2$   & 3.2414 & 0.2943 \\
    $p=0.25$  & 3.1773 & 0.2841\\
    $p=0.3$  & \textbf{3.0651} & \textbf{0.2734} \\
    $p=0.35$  & 3.1990 & 0.2909 \\
    \bottomrule
  \end{tabular}}
  \label{tab:droppath}
\end{table}

\subsubsection{PEG}

\textbf{Table \ref{tab:peg}} reports the performance of PoFormers with different positional encoding strategies. Either removing the PEG or replacing it with a sinusoidal positional encoding degrades its performance. Sinusoidal encodings causes the performance to decay because the test audio segments vary in length while the training sequences have the same length. Moreover, as long as PEG is used, the result is insensitive to the kernel size as PEGs with different kernel sizes give similar EER and minDCF.

\begin{table}[htbp]
  \centering
  \caption{Ablation study on PEG. PoFormers trained with different PEGs and evaluated on Voxceleb1-H are presented.}
  \label{tab:tabel_three}
  \setlength{\tabcolsep}{6mm}{
  \begin{tabular}{lcc}
    \toprule
    {PEG kernel size $k$} & {\textbf{EER(\%)}} & {$\textbf{DCF}_\textbf{0.01}$} \\
    \midrule
    No PEG  & 3.0651 & 0.2734 \\
    sinusoidal & 3.3023 & 0.2951 \\
    $k=3$   & 2.9384 & 0.2625 \\
    $k=9$   & 2.8754 & \textbf{0.2572} \\
    $k=15$  & \textbf{2.8555} & 0.2654 \\
    \bottomrule
  \end{tabular}}
  \label{tab:peg}
\end{table}

\subsubsection{Pre-norm and post-norm}

\begin{figure}[ht]
  \centering
  \includegraphics[width=0.3\textwidth]{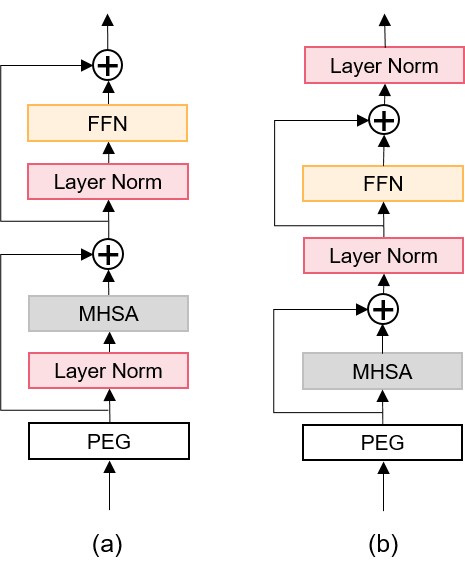}
  \caption{The difference between (a) pre-norm transformer and (b) post-norm transformer.}
  \label{fig:pre_post}
\end{figure}

As is mentioned in section \ref{sec:methods}, our PoFormer uses the pre-norm transformer \cite{xiong2020on}, where the layer normalization is applied before the MHSA module and the FFN module. However, the original transformer structure in \cite{google_2017_attention} is a post-norm transformer. Their difference is illustrated in \textbf{Fig.\,\ref{fig:pre_post}}. To figure out which one is more suitable to speaker verification, we compare their performances, and the results are listed in \textbf{Table \ref{tab:norm}}. We can see that the pre-norm PoFormer outperforms the post-norm one.

 \begin{table}[htbp]
  \centering
  \caption{Ablation study on pre-norm and post-norm. PoFormers with pre-norm and post-norm transformer layer are trained. Both are evaluated on Voxceleb1-H.}
  \label{tab:tabel_three}
  \setlength{\tabcolsep}{7.8mm}{
  \begin{tabular}{lcc}
    \toprule
     & {\textbf{EER(\%)}} & {$\textbf{DCF}_\textbf{0.01}$} \\
    \midrule
    pre-norm     & \textbf{2.8754} & \textbf{0.2572} \\
    post-norm    & 2.9251 & 0.2695 \\
    \bottomrule
  \end{tabular}}
  \label{tab:norm}
\end{table}

\subsubsection{Class token}

In our standard PoFormer, we use an extra class token to embed and classify the speaker. However, we find that concatenating the mean and the standard deviation of the frame-level output from the last transformer layer provides a slightly better performance, which is a bonus for our PoFormer. The results are reported in \textbf{Table \ref{tab:cls}}.

\begin{table}[htbp]
  \centering
  \caption{Ablation study on class token. PoFormer with extra statistic information provides a slightly better performance on Voxceleb1-H.}
  \label{tab:tabel_three}
  \setlength{\tabcolsep}{6mm}{
  \begin{tabular}{lcc}
    \toprule
    {PoFormer output} & {\textbf{EER(\%)}} & {$\textbf{DCF}_\textbf{0.01}$} \\
    \midrule
    class token      & 2.8754 & 0.2572 \\
    class token + stats   & \textbf{2.8295} & \textbf{0.2495} \\
    \bottomrule
  \end{tabular}}
  \label{tab:cls}
\end{table}

\section{Conclusion}
\label{sec:conslusion}
In this paper, we propose a transformer-based pooling structure, PoFormer for the speaker verification systems. To the best of our knowledge, our work is the first attempt to introduce the original transformer structure into the pooling layer. The multi-head self-attention mechanism can effectively capture global information. By adding PEG, LayerScale and drop path, our PoFormer outperforms all the existing pooling system in speaker verification.

\vfill\pagebreak

\label{sec:refs}



\end{document}